\documentclass[prb,twocolumn,superscriptaddress,showpacs]{revtex4}
\usepackage{mathrsfs}
\usepackage{amsmath,amsfonts,amssymb}
\usepackage{epsfig}
\usepackage{graphicx}
\usepackage{wasysym}
\usepackage{bbm}
\usepackage{psfrag}
\usepackage{color}
\usepackage{pstool}
\usepackage{braket}
\usepackage[dvips, bookmarks, colorlinks=true, plainpages = false, citecolor = blue, linkcolor = blue, urlcolor = blue, filecolor = blue]{hyperref}
\newcommand \be{\begin{equation}}
\newcommand \ee{\end{equation}}
\newcommand \bes{\begin{equation*}} 
\newcommand \ees{\end{equation*}}
\newcommand \bea{\begin{eqnarray}}
\newcommand \eea{\end{eqnarray}}
\newcommand \bsea{\begin{subequations}\begin{eqnarray}} 
\newcommand \esea{\end{eqnarray}\end{subequations}}
\newcommand \beas{\begin{eqnarray*}} 
\newcommand \eeas{\end{eqnarray*}}
\newcommand \bfg{\begin{figure}}
\newcommand \efg{\end{figure}}
\newcommand \bfgs{\begin{figure*}} 
\newcommand \efgs{\end{figure*}}
\newcommand \bwt{\begin{widetext}}
\newcommand \ewt{\end{widetext}}

\def\pmat#1{\left(\begin{matrix}#1\end{matrix}\right)}

\begin{document}
\title{Nontrivial Topological Phases on the Stuffed Honeycomb Lattice }
\author{Arghya Sil}\email{arghyasil36@gmail.com}
\affiliation{Department of Physics, Jadavpur University, 188 Raja Subodh Chandra Mallik Road, Kolkata 700032, India}
\author{Asim Kumar Ghosh}\email{asimkumar96@yahoo.com}
\affiliation{Department of Physics, Jadavpur University, 188 Raja Subodh Chandra Mallik Road, Kolkata 700032, India}

\begin{abstract}
We report the appearance of nontrivial topological phases in a 
tight-binding model on the stuffed honeycomb lattice. The model contains 
nearest neighbor and next nearest neighbor hopping terms 
coupled with an additional phase
depending on the direction of hopping.  
Chern insulating and semi-metallic phases emerge with the  
change of hopping parameters. Nonzero Chern numbers characterizing the bands 
and the existence of topologically protected edge states in the gap between the
relevant bands confirm the presence of those phases. 
We show that adding an extra basis to Haldane's 
honeycomb model can lead to an additional topological phase characterized
by Chern number $\pm2$. Transition between different topological phases
driven by the hopping parameters has been illustrated 
in the topological phase diagram of the system.
Zero temperature Hall conductivity along with density of states is 
evaluated. Topological properties of another tight-binding model on 
the stuffed square lattice are also reported in this article. 
\end{abstract}
\maketitle

\section{Introduction}
Investigation of nontrivial topological phases on various 
two-dimensional lattice 
models has been increasing rapidly in the recent times. 
These studies are motivated by the search of novel topological
phases within new tight-binding models. 
These models are primarily characterized by a set of special 
energy bands those are separated by bulk energy gap but
joined by quasi-continuous edge state energies.  
These are known as topological
insulators (TI) which are distinguished by a  
topologically invariant integral number such as  
Chern number ($C_n$) \cite{TKNN}, $Z_2$ invariant, \cite{Kane1} etc.

It is established that the emergence of topological phases 
is the handiwork of specific phase factors which are incorporated  
into the tight-binding models through the hopping terms. 
In case of multipartite lattices, this phase factor 
appearing in the momentum space representation of the Hamiltonian  
is coupled with the hopping term between same or different sublattices.  
For instance, it arises in the one-dimensional bipartite 
Su-Schrieffer-Hegger model \cite{SSH} only through the 
nearest neighbor (NN) hopping between two different sublattices. 
In this case, it depends on the Bloch wave vector, $\textbf{k}$.  
For the two-dimensional bipartite honeycomb lattice, 
an additional phase coupled with the next-nearest-neighbor (NNN) 
hopping between the same sublattice plays the crucial role 
to generate the nontrivial topological phases \cite{Haldane}.
Here, the sign of this phase is  opposite 
for the two different sublattices. 
The net flux passing through a unit cell 
due to this gauge field is zero.  
The resulting Hamiltonians in these two cases  
break the Time Reversal Symmetry (TRS) due to the presence of this phase factor 
for which the topological energy bands are characterized by $C_n$. 
In other words, this phase factor drives the 
system into an Integer Quantum Hall Effect (IQHE) regime.  

On the other hand, $Z_2$ topological phase appears in the 
honeycomb lattice due to the presence of spin-orbit coupling (SOC) 
for which the resulting Hamiltonian does not break the TRS \cite{Kane1}. 
Here, the system is characterized by $Z_2$ invariant. 
However, if the Hamiltonian commutes with the $z$-component of spin 
operator, $S_z$, then the Hamiltonians for up and down spins break 
the TRS separately. In this case, resulting energy bands  
are characterized by spin-Chern numbers. 

For a trivial insulator, the topological invariant is zero, while for a
nontrivial TI, it can assume nonzero values.
For example, there is nonzero $C_n$ for Chern insulators.  
Also, the number of edge states is proportional to the value of $C_n$,  
which is known as the `bulk-boundary 
correspondence' \cite{Hatsugai1,Hatsugai2,Kane2}. Due to this correspondence, the edge states 
which appear in a strip geometry become topologically protected.   
Thus, topological invariants are helpful for the classification of 
the topological phases of a system. 
A brief review on various TIs is available in the article \cite{Das}.

Although the Chern insulating phase appears for TRS breaking Hamiltonians, 
the converse is not true. We cannot readily obtain nontrivial topological phase
for all TRS breaking Hamiltonians. So, several attempts 
have been made in search of this phase in various multi-band 
systems. A number of two-dimensional lattices with nontrivial topological 
phase have been found and they are  
Lieb \cite{Franz}, kagome \cite{Wen}, checkerboard 
\cite{Das-sharma1}, square octagon \cite{Kargarian}, dice\cite{Wang,Gong2}, 
star lattice \cite{Chen}, etc. 
In addition, different topological flat-band models have 
also been proposed, where Fractional Quantum Hall Effect (FQHE) 
can be realized as  
those flat-bands carry nonzero Chern number \cite{Das-sharma2}. 

Recently, nontrivial bands have been realized in optical lattices of 
ultracold atoms \cite{Wu} by tuning the strengths of both NN 
and NNN hopping amplitudes \cite{Weiss} 
and thus realization of artificial gauge field has been 
made possible\cite{Spielman,Aidelsburger}. 
This has opened a new path of obtaining 
Chern insulator by varying the hopping parameters of the model
tight-binding Hamiltonian in various two-dimensional 
lattice-systems \cite{Bergholtz,Gong1,Beugeling}.

In this work, we search for nontrivial topological phase
on the stuffed honeycomb lattice, which is a three-band system by itself 
but actually interpolates 
the triangular (Bravais) 
and two-band honeycomb (non-Bravais) lattices. 
Being a two-band system honeycomb model is capable to 
exhibit nontrivial topological phase in the presence 
of phase-coupled hopping terms,  
while triangular lattice fails to do so under the same 
situation for obvious reason. 
Therefore, emergence of new topological phase in the 
interpolating regime will be of great interest. 
Eventually it turns out that addition of extra sublattice 
does lead to the emergence of new topological 
phase in the resulting system. In support of this claim, 
another tight-binding model on stuffed square lattice is 
hereby introduced which is capable to host 
new topological phase.

The antiferromagnetic (AFM) Heisenberg model on this lattice was studied previously 
in search of new spin-liquid phase. However, in that case, 
triangular, honeycomb as well as the intermediate stuffed honeycomb 
lattices were found to host similar kind of spin-liquid phase \cite{Sahoo}.
Interpolating lattices are capable to exhibit nontrivial spin chirality, 
which may lead to anomalous Hall Effect \cite{Nagaosa}. S
o, adding another basis to Haldane's honeycomb model
could provide the platform to explore novel topological phases in tight-binding regime. 
With this motivation,
we formulated a three-band tight-binding model on this lattice in the presence of NN and NNN 
hopping terms. An additional phase coupled with NNN hopping  
has been considered to incorporate nontriviality in the 
otherwise trivial system. The sign of this phase 
depends on the direction of hopping. This phase breaks the TRS in the 
momentum space representation of the Hamiltonian. 
Also, the net flux of the gauge field passing through a unit cell 
is zero. The system is shown to exhibit distinct 
TI phases characterized by different sets of Chern numbers.
Transition between those phases are driven by the modulation of hopping strengths. 
Chiral edge states are simultaneously obtained in the finite system.
 
The plan of the paper is as 
follows. In section \ref{model}, stuffed honeycomb lattice is 
described and the tight-binding Hamiltonian is formulated. 
This is followed by section \ref{properties}, where behavior of the relevant physical 
quantities to study 
topological properties is presented. 
Finally, in section \ref{summary}, discussions and conclusions on this work are summarized.
   
\section{Stuffed Honeycomb Lattice and Formulation of Hamiltonian}
\label{model}
\begin{figure}[t]
\includegraphics[width=8cm]{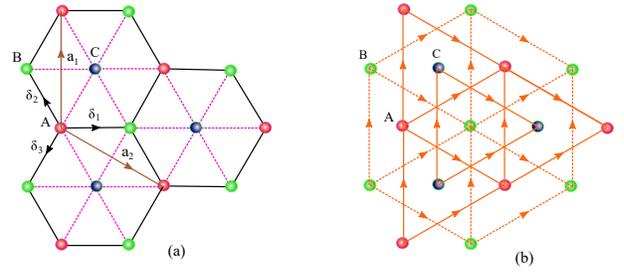}
\caption{(color online) (a) NN interactions are shown. Hopping amplitude is $t_1$ 
along the solid lines and $t_2$ along dashed lines. 
Lattice vectors are shown as $\mathbf{a_1}$ and 
$\mathbf{a_2}$. Three types of lattice sites
$A, B, C$ are drawn as red, green and blue spheres, respectively.
$\mathbf{\delta_1},\mathbf{\delta_2}, 
\mathbf{\delta_3}$ are the three NN vectors. 
(b) NNN interactions are shown. Hopping amplitude is $t_3$ along the solid 
lines and $-t_3$ along dashed lines. Sign of the phase 
is assumed positive when hopping is along the direction of arrow 
and otherwise negative.}
\label{lattice}
\end{figure}

Stuffed honeycomb lattice originates as a result of coupling 
between one honeycomb lattice and one triangular lattice 
which is shown in Fig. \ref{lattice}(a). The honeycomb lattice is 
composed of two triangular lattices with site indices $A$ and $B$, while the 
site index of the additional triangular lattice is $C$.  So, essentially 
it is tripartite and composed of three interpenetrating identical triangular lattices. 
The resulting lattice is decomposed in this manner to 
define the hopping parameters for this 
tight-binding model in a comfortable way. 
We consider three different species of spinless fermions each for three 
sublattices to introduce the Hamiltonian
\bea
 H\!\!&=&\!\! -\sum_{\langle ij\rangle} \left(t_1 \;A_i^{\dagger}B_j +t_2 \;(A_i^{\dagger}C_j+B_i^{\dagger}C_j)+H.c \right) \nonumber\\
 &-&\sum_{\langle\langle ij\rangle\rangle} \left(t_{ij}\; e^{i\phi_{ij}}\; ({A_i}^{\dagger}A_j+{B_i}^{\dagger}B_j+{C_i}^{\dagger}C_j) +H.c \right) \nonumber\\
 &+& \sum_i \epsilon_i\left({A_i}^{\dagger}A_i+{B_i}^{\dagger}B_i+{C_i}^{\dagger}C_i  \right),
\eea
where $i$ is the site index. $\langle \cdot\rangle$ and 
$\langle\langle \cdot\rangle\rangle$ indicate the summations over 
NN and NNN pairs.  ${\alpha_i}^{\dagger}(\alpha=A,B,C)$ is 
the fermion creation operator at site $i$.
$t_1$ is the intra-honeycomb NN hopping amplitude and $t_2$ 
is the NN hopping amplitude between the honeycomb and triangular sublattices 
(Fig. \ref{lattice}(a)). 
$t_{ij}$ is the NNN hopping amplitude irrespective of  
sublattices but $t_{ij}$ = $t_3\, (-t_3)$ for solid (dashed) 
lines as shown in Fig. \ref{lattice}(b).
The direction of the phases $\phi_{ij}=\pi/2$ is indicated 
by the arrow in Fig. \ref{lattice}(b).
$\epsilon_i$ is the onsite energy. 

This model will become the honeycomb one when both 
$t_2$ and  $t_3$  across the $C$-$C$ bonds vanish. 
Similarly, it interpolates separately to 
triangular and dice lattices at $t_2$=$t_1$ and 
at $t_2$=$\infty$, respectively \cite{Sahoo}. The unit cell is defined by the 
primitive vectors  $ \mathbf{a_1}= \sqrt3\left(0,1\right)$ and 
$\mathbf{a_2} = \sqrt3\left(\sqrt{3}/2,-1/2 \right)$, 
where the NN distance is taken to be unity.
NN sites are connected by the vectors $\mathbf{\delta_1}, 
\mathbf{\delta_2}$ and 
$\mathbf{\delta_3}$ where $\mathbf{\delta_1} = \left(1,0\right) 
= 1/3 \left(\mathbf{a_1}  +2 \mathbf{a_2}  \right),\; 
\mathbf{\delta_2}
= \left(-1/2,\sqrt{3}/2\right) = 1/3 \left(\mathbf{a_1}  -
 \mathbf{a_2} \right)$ and 
$\mathbf{\delta_3} = \left(-1/2,-\sqrt{3}/2\right) = 
-1/3 \left( 2\mathbf{a_1}  +\mathbf{a_2} \right) $. 
All those vectors are shown in 
Fig. \ref{lattice}(a). 
The corresponding reciprocal lattice vectors are 
$\mathbf{b_1}=\left(2\pi/3,2\pi/\sqrt{3} \right)$ and 
$\mathbf{b_2}=\left(4\pi/3,0 \right)$ which span the hexagonal
first Brillouin zone.
 
Hamiltonian in the momentum space is written by invoking periodic 
boundary conditions along both $\mathbf{a_1}$ and $\mathbf{a_2}$ 
directions,  
\be H\left(\textbf{k}\right)=\sum_k {\psi_k}^{\dagger} h\left(\textbf{k}
\right) \psi_k, \ee
 where $\textbf{k} =(k_x,k_y)$, $\psi_{\textbf{k}} = \left(A_{\textbf{k}},
B_{\textbf{k}},C_{\textbf{k}}\right) $ is a $3$-component 
spinor and $h(\textbf{k})$ is a $3\times 3$ matrix.
 This $h(\textbf{k})$ can be expressed in 
terms of eight Gell-Mann matrices,  
$\lambda_i$ as \be  h\left(\textbf{k}\right) = \sum_i h_i\, \lambda_i + a\, I_3,\ee
 with
 \begin{equation}
 \begin{aligned}
 h_1&=t_1\big[\cos\left(\frac{k_1+2k_2}{3}\right)+\cos\left(\frac{k_1-k_2}{3}\right) \\
      &+\cos\left(\frac{2k_1+k_2}{3}\right)\big], \\
 h_2&=-t_1\big[\sin\left(\frac{k_1+2k_2}{3}\right)+\sin\left(\frac{k_1-k_2}{3}\right)\\
     &-\sin\left(\frac{2k_1+k_2}{3}\right)\big], \\
 h_3&=2t_3f(k_1,k_2), \\
 h_4&=h_1\left(\frac{t_2}{t_1}\right) = h_6, \\
 h_5&=-h_2\left(\frac{t_2}{t_1}\right) = -h_7, \\
h_8&=-\frac{4}{2\sqrt{3}}t_3f(k_1,k_2) +\sqrt{3}\epsilon, \\
 a&=\frac{2}{3}t_3f(k_1,k_2),
 \end{aligned}
\end{equation}
where $I_3$ is the $3\times 3$ identity matrix,  $k_1=\textbf{k} \cdot \mathbf{a_1} = \sqrt{3} k_y $,  
$k_2=\textbf{k} \cdot \mathbf{a_2} = 3/2 k_x - \sqrt{3}/2 k_y $
and $f(k_1,k_2)=\big[\cos\left(k_1+\pi/2\right)+\cos\left(k_2+\pi/2\right)
      +\cos\left(k_1+k_2+\pi/2\right)\big]$. 
$\lambda_i$ are shown in the Appendix  \ref{gm}.  
We have written the Hamiltonian matrix in terms of $k_1$ and $k_2$
for ease of calculation.
Evidently, TRS is broken by the phase
$\phi_{ij}=\pi/2$ since $h(\textbf{k})\neq h^{\ast}(-\textbf{k})$.  
Hopping parameters $t_2$ and $t_3$ as well as $\epsilon_i$ 
are measured with respect to $t_1$ which is taken as $1.0$ throughout the article. 
The value of $\epsilon_i$ is taken as
$\epsilon$ ($-2\epsilon$) for honeycomb (triangular) sublattice. 
This special distribution of values of $\epsilon_i$ 
is necessary to open up gaps between the otherwise gapless energy bands. 

 Signs of hopping terms are taken in such a way that 
one among the three terms $h_n (n=2,5,7)$ has 
sign opposite to that of the remaining two terms. This reverses
the bloch phase of one of the off-diagonal elements of the 
Hamiltonian with respect to the other two (see Appendix  \ref{gm}).
We note that in the case of SU(3)-invariant models such a 
situation is a necessary criterion for generating the 
nontrivial topology \cite{Das2}.

\begin{figure}[h]
  \psfrag{Kx}{$k_x$}
\psfrag{Ky}{$k_y$}
\psfrag{c0}{\text{\scriptsize{\bf \color {white} {$C_n=0$}}}}
\psfrag{c2}{\text{\scriptsize{\bf \color {white} {$C_n=2$}}}}
\psfrag{c-2}{\text{\scriptsize{\bf \color {white} {$C_n=-2$}}}}
\psfrag{c1}{\text{\scriptsize{\bf \color {white} {$C_n=-1$}}}}
\psfrag{c-1}{\text{\scriptsize{\bf \color {white} {$C_n=1$}}}}
  \includegraphics[width=4cm,height=6cm,trim={2.0cm 0.0cm 3.5cm 0.0cm}]{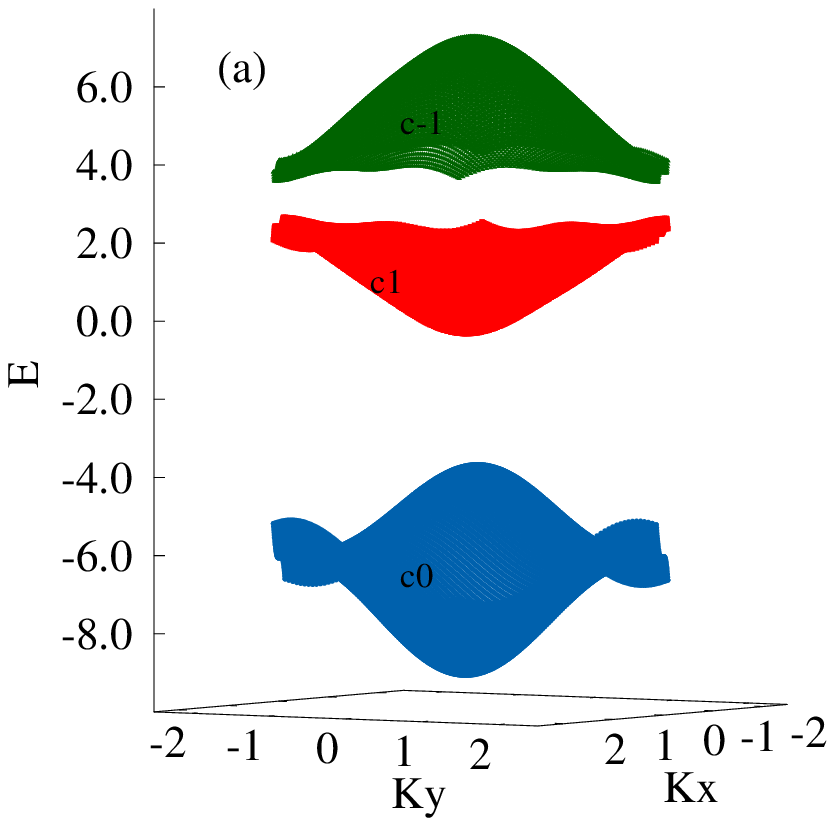}
  \includegraphics[width=4cm,height=6cm,trim={1.5cm 0.0cm 3.5cm 0.0cm}]{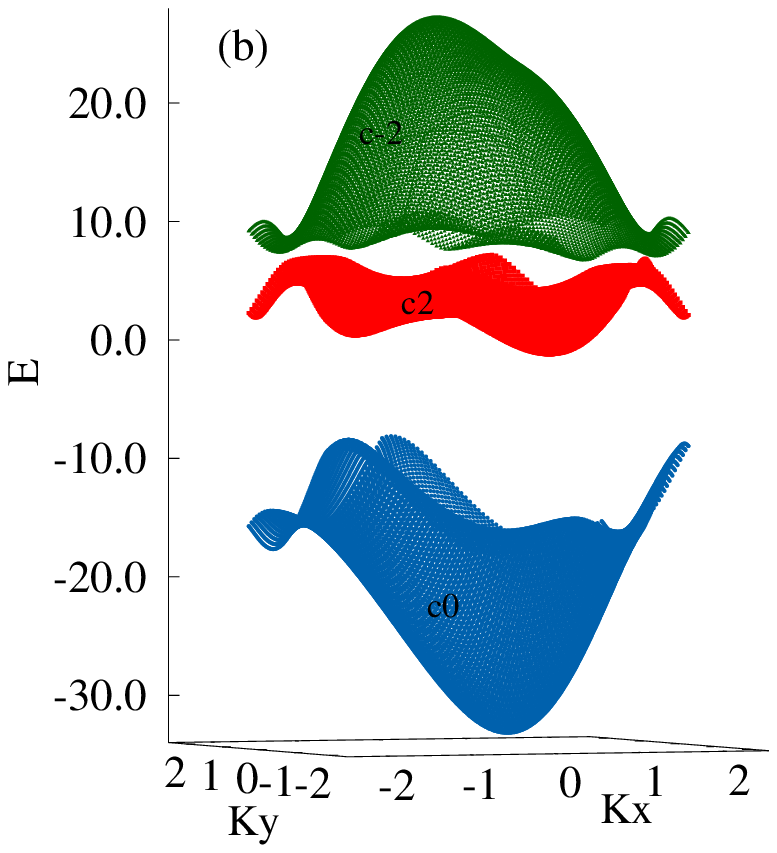}
   \caption{(color online) Dispersion relation for (a) $t_2$=$0.8$, $t_3$=$0.5$, 
$\epsilon$=$3.0$ and (b) $t_2$=$6.0$, $t_3$=$2.0$, $\epsilon$=$6.0$. 
$t_1$ is taken to be unity. 
The Chern numbers of the respective bands are specified.}
 \label{dispersion}
  \end{figure}

Analytic expressions for the eigenvalues of $h(\textbf{k})$ are 
given by 
\be E_m(\textbf{k})=2\sqrt{\frac{-p}{3}}\cos{\left(\frac{1}{3}\arccos{\left(\frac{3q}{2p}
\sqrt{\frac{-3}{p}}\right)}-\frac{2\pi m}{3}\right)},\ee
 $m=0,1,2$;
where the real parameters are 
\begin{equation}
 \begin{aligned}
p&= -\frac{4}{3}h_{3}^{2}(\textbf{k})+2h_{3}\,\epsilon-({t_1}^{2}+2{t_2}^{2}){|f(\textbf{k})|}^{2}-3\,{\epsilon}^{2}, \\
 q&=\frac{16}{27}\,{h_{3}}^{3}(\textbf{k})
   -\frac{4}{3}\,{h_{3}}^{2}(\textbf{k})\,\epsilon
   -t_{1}{t_2}^{2}(f^{3}(\textbf{k})+{f^{\ast}}^{3}(\textbf{k}))\\
   &-2({t_1}^{2}-{t_2}^{2})\left(\frac{h_{3}(\textbf{k})}{3}-2\,\epsilon\right){|f(\textbf{k})|}^{2}
   .
 \end{aligned}
\end{equation}
Here, $f(\textbf{k})=(h_{1}(\textbf{k})-ih_{2}(\textbf{k}))/t_1$.
The dispersion relations are shown in Fig. \ref{dispersion} (a) and (b) 
for two sets of $(t_2,t_3,\epsilon)$ where nonzero Chern numbers are found.

\section{Topological Properties}
\label{properties}
In order to study the topological properties, Chern numbers, 
Hall conductance at zero temperature $(\sigma_H)$ and  
edge states of this model have been calculated. 
\subsection{Chern Numbers and Topological Phase Transition}
In the beginning, we calculate $C_n$ of the three bands 
to characterize the topological phases of this system. 
$C_n$ is defined as the integration of the Berry curvature 
$\Omega_n (\textbf{k})$ over the first Brillouin zone (1BZ), {\em i.e.}, 
\begin{equation}
 \begin{aligned}
  C_n &=\frac{1}{2\pi}\int_{1BZ}d^2\textbf{k} \cdot \Omega_n\left(\textbf{k} \right),
    \label{Cn}
 \end{aligned}
\end{equation}
where $\Omega_n\left(\textbf{k} \right)=
-i\left(\braket{\partial_{1} u_{n,\textbf{k}}|\partial_{2} u_{n,\textbf{k}}}-
    \braket{\partial_{2} u_{n,\textbf{k}}|\partial_{1} u_{n,\textbf{k}}}\right)$. 
Here $\ket{u_{n,\textbf{k}}}$ are the eigenvectors of $H(\textbf{k})$ 
and $\partial_{i}=\frac{\partial}{\partial k_i}$. 
$C_n$ is well-defined for a particular band as long as it does not 
touch other neighboring bands {\em i.e.}, the eigenvalues $E_n(\textbf{k})$ 
are not degenerate for any fixed $\textbf{k}$. 
In our numerical calculation, we use the discretized version of 
Eq \ref{Cn} introduced by Fukui and others \cite{Suzuki}. 

\begin{figure}
\centering
  \includegraphics[width=8cm,trim={0.0cm 0.0cm 0.0cm 0.0cm}]{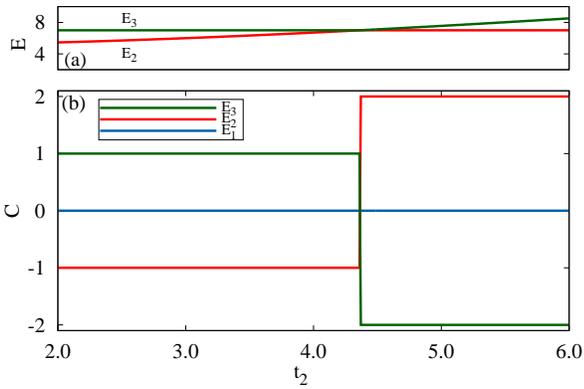}
 \caption{(color online) 
 (a) Plot of two upper energies (red and green lines) with respect to 
$t_2$ showing the crossing between them 
 for $(k_x,k_y)=(\pi,0)$ or $(0,\pi)$ or $(\pi,\pi)$, (same value in each case), 
$t_3=2.0$ and $\epsilon=6.0$. The lowest energy (blue line) is not shown here.
(b) Topological phase diagram for $t_3=2.0$ and $\epsilon=6.0$ with varying $t_2$.
 The energy bands are denoted by $E_n(n=1,2,3)$ in ascending order of energy.
 Summation of Chern numbers over all the bands are zero, which is obvious in the figure. 
By comparing the diagrams (a) and (b), it is evident that upper band gap closes at 
the phase transition point, $t_2=4.35$.}
 \label{phase}
\end{figure}

\begin{figure*}
\centering
  \includegraphics[width=16cm,height=5.5cm]{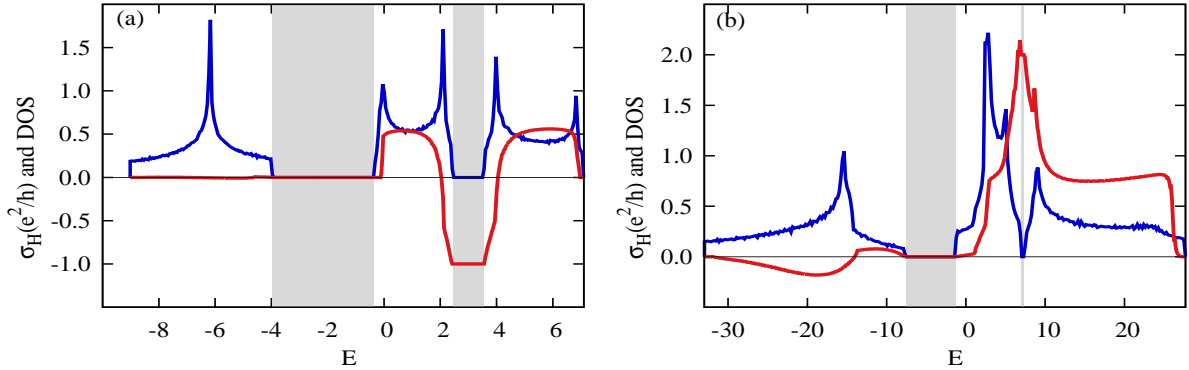}
 \caption{(color online) The Hall conductance $\sigma_H$ (red line) and DOS (blue line) 
with respect to the Fermi energy $E$ for (a) $t_2=0.8$, $t_3=0.5,\,\epsilon=3.0$ and
 (b) $t_2=6.0,\,t_3=2.0,\,\epsilon=6.0$. The shaded regions are showing the 
band-gaps.}
 \label{hall}
\end{figure*} 

In this model, the presence of TRS breaking NNN phase, $\phi_{ij}$ 
gives rise to the non-vanishing Chern number, since the 
TRS invariant Hamiltonian leads to 
$\Omega_n(\textbf{k})=-\Omega_n(-\textbf{k}) $ resulting to 
$C_n=0$.  Evolution of both the dispersion relation and the  
 topological phases is studied in a space spanned by 
the three parameters $t_2$, $t_3$ and $\epsilon$. $C_n$ does not change its sign if the sign of $t_2$ is reversed. 
However,  $C_n$ is found to change its sign when either
$t_3$ or $\phi_{ij}$ changes its sign. 

Let us now explain the evolution of topological phases along definite lines 
in this three-dimensional parameter space. 
At first, $\epsilon$ and $t_3$ are kept fixed at 6.0 and 2.0, respectively,  
while $t_2$ is allowed to vary.   
For $t_2=0$, the two lower bands touch each other so that Chern 
numbers are ill-defined. As soon as $t_2$ becomes greater than $0.0$, 
a pseudo-gap is opened up between the two lower bands and it remains so 
upto $t_2<1.3$. This pseudo-gap becomes a true gap when  
$t_2>1.3$. Along this line, there is always a true 
gap between the two upper bands as long as $0.0<t_2< 4.3$. 
Chern numbers are defined in these pseudo-gapped regions because of 
the fact that although the Fermi energy partially crosses 
one or more bands, no bands are found to touch each other. 
On the other hand, Fermi energy never crosses 
any band where true band gap exists. 
Pseudo-gapped phase is known as Chern semi-metallic phase, 
whereas the true gapped phase is known as Chern insulating phase.
Thus, nontrivial topological phases appear over this line 
when $0.0<t_2<4.3$, which is characterized by the value of $C_n=(0,-1,1)$. 
Two upper bands touch each other at $t_2=4.35$ 
where the system undergoes a topological phase transition. 
The closing of band gap at the transition point is essential 
to ensure the topological phase transition \cite{Kane2}. 
When $t_2>4.4$, the Chern numbers are redistributed as $C_n=(0,2,-2)$.
The Chern number exchange of $\Delta C_n = \pm 3$ may attribute to the fact
that at $t_2=4.35$, two upper bands touch each other simultaneously at
three different points in the first Brillouin zone, say, 
$(k_x,k_y)=(0,\pi),(\pi,0),(\pi,\pi)$, each causing an
exchange of $\Delta C_n=\pm 1$.
No new topological phase is found to emerge upon further 
increase of $t_2$, since no band crossing is observed.  

In the same way, let us explore the evolution of topological phases 
along another line by keeping $t_2$ and $\epsilon$ 
fixed at $0.5$ and $3.0$, respectively. 
At $t_3=0.0$, the upper two bands touch each other and the 
Chern numbers are undefined.
As soon as $t_3$ becomes nonzero, a nontrivial phase with $C_n=(0,-1,1)$ 
appears but the system exhibits true band gaps upto $t_3=1.0$. 
Afterwards, a pseudo gap develops between the two lower bands.
This semi-metallic phase persists upto $t_3=2.0$. 
For $t_3>2.0$, the upper band-gap closes and the system becomes trivial.
No new topological phase is found to appear further along this line. 

Similarly, examining along many other lines 
in the parameter space, no new phase other than 
these two topologically different phases 
characterized by $C_n=(0,-1,1),\,(0,2,-2)$ are found to 
exist. In every case, Chern number for the lowest band is 
zero. The phase diagram is presented in Fig \ref{phase} 
for $t_3=2.0$ and $\epsilon=6.0$ indicating transitions between different phases. 
It is evident that these topological phases are robust 
against change of external parameters as long as the alteration  
does not cause another band-touching or gap-closing.

The lowest energy band is partially
localized on the triangular sublattice because of relatively strong 
sublattice potential, $-2\epsilon$, with respect to 
that of honeycomb sublattice, $\epsilon$. 
In the honeycomb limit, ({\em i. e.} $t_2=0$ and across
C-C bond $t_3=0$), we get only a single phase, $C_n=\pm 1$. 
Additional $C_n=\mp 2$ phase 
appears after turning on $t_2$ which results the stuffed honeycomb lattice. 
This fact may be termed as generation of additional 
topological phase due to the entry of an additional sublattice.

It would be worth mentioning in this context that a hard-core bosonic model 
based on this stuffed-honeycomb lattice has been studied before, where   
the topological phase with all topologically 
non-trivial bands $C_n=(-1,-1,+2$) is found. 
Additionally, the lowest band ($C_n=2$) 
bears high flatness ratio, 15, which gives rise to bosonic FQHE 
at 1/3 filling and fermionic FQHE at 1/5 filling \cite{Wang1}. 
On the other hand, in our fermionic model, 
although the lowest band becomes nearly flat 
when the values of both $t_2$ and $t_3$ are very small, the system 
is not a potential candidate for FQHE states 
as this band always carries zero Chern number. 
However, the topological phase ($C_n=-1,-1,+2$) observed in the bosonic model 
could be realized in the fermionic counterpart by mimicking both the 
hopping terms and phase values like the former. As well as, 
FQHE state could be found in the lowest non-trivial flat band.
\subsection{Hall conductance at zero temperature}
At zero temperature, $\sigma_H(E)$ is
 estimated numerically by using the Kubo formula \cite{TKNN} 
\begin{align*} 
\sigma_H(E)&=\frac{ie^2\hbar}{A_0}\sum_{\textbf{k}}\sum_{E_m<E<E_n}\nonumber\\
 &\frac{\langle m|v_x|n\rangle 
\langle n|v_y|m\rangle\!-\!\langle m|v_y|n\rangle 
\langle n|v_x|m\rangle}{(E_m-E_n)^2},
\end{align*}

where $\ket{l}=\ket{u_{l,\textbf{k}}}, H_{\textbf{k}}\ket{l}=
E_l\ket{l}$ and $l=m,n$. $A_0$ is the area of the system and 
$E$ is the Fermi energy. The velocity operator, 
$v_\alpha=(1/i\hbar)[\alpha,H]$, where $\alpha=x,y$. 
When $E$ falls in one of the energy gaps, 
the contribution to $\sigma_H$
by $n$ completely filled bands is given by 
\be 
\sigma_H\left(E\right)=\frac{e^2}{h}\sum_{E_n<E}C_n. 
\label{hall-plateau}
\ee 
In this situation, $\sigma_H(E)$ always assumes an integral value. 

 $\sigma_H(E)$ along with the density of states (DOS) is 
plotted against Fermi energy in Fig \ref{hall} 
for two different sets of parameters where 
two distinct topological phases are observed. 
DOS is useful to locate the gap in the band diagram 
where $\sigma_H(E)$ always exhibits a Hall plateau. 
The height of a Hall plateau can be determined by 
using Eq \ref{hall-plateau}. 
Two prominent Hall plateaus for $\sigma_H=n(e^2/h)$ with $n=(0,-1)$ 
are observed in Fig \ref{hall}(a), 
which corresponds to the topological phase 
having $C_n=(0,-1,1)$.  
Similarly, in Fig \ref{hall}(b) two Hall plateaus exist for 
$\sigma_H=n(e^2/h)$ with $n=(0,2)$, but the second one is not prominent 
in the figure since the band gap is very narrow in this case. 
This characteristic corresponds to the 
topological phase having $C_n=(0,2,-2)$.  
Band gaps are identified by the shaded regions 
which hold these Hall plateaus. 
According to Eq \ref{hall-plateau}, the value of $\sigma_H(E)$ over any  
Hall plateau becomes equal to the sum of all Chern numbers carried by 
the bands having energy lower than it. 
So, the sign of $\sigma_H(E)$ be either positive or negative 
depending on the distribution of $C_n$ over the energy bands. 
Also, width of the band gap equals to that of the plateau. 
DOS exhibits sharp peaks around the energies where 
$\sigma_H(E)$ undergoes sharp rise and fall. As a result, 
four sharp peaks in DOS are found in Fig \ref{hall}(a). 
Another sharp peak in DOS at the lowest energy 
corresponds to the van Hove singularity in the band diagram.

\subsection{Edge States}
\begin{figure*}
\centering
  \includegraphics[width=16cm,height=7cm,trim={0.0cm 0.2cm 0.0cm 0.0cm}]{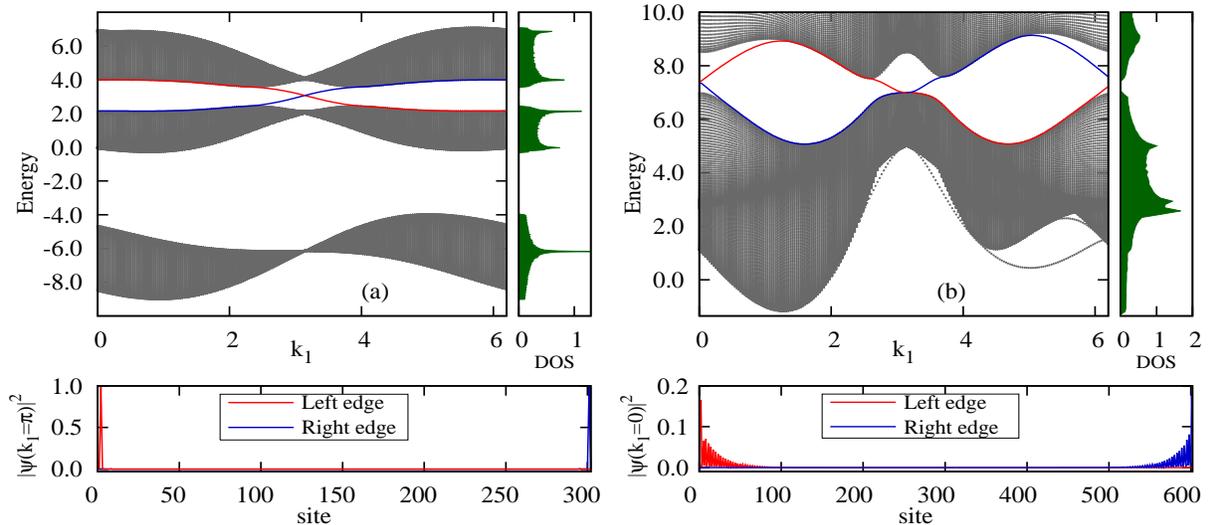}
 \caption{ (color online) Edge states of the stuffed honeycomb lattice 
considering zigzag edge along 
$\mathbf{a_2}$ direction are shown in solid lines 
for (a) $t_2=0.8,\,t_3=0.5,\,\epsilon=3.0$
  with 100 cells and (b) $t_2=6.0,\,t_3=2.0,\,\epsilon=6.0$ with 
200 cells. For (b), 
only the upper band-gap is shown as the energy range being 
 very high, lower gap containing zero edge states is not shown. 
Lower panels demonstrate the distribution of probability densities 
of a particular eigenstate with $k_1=\pi$ for (a) and $k_1=0$ 
for (b). Red (blue) line corresponds 
to left (right) edge. }
 \label{edge}
\end{figure*}

Among the topological properties, Chern number is recognized as the bulk 
property of the system, where edge states correspond  
to the surface property. However, the presence of non-zero Chern number
leaves its signature by generating edge states, and vice versa. To study the 
properties of edge states in case of nontrivial Chern number, 
boundary lines or edges  are created by removing the periodic boundary 
condition along $\mathbf{a_2}$ axis, such that, $k_2$ is no longer  
a good quantum number for this finite system. But the  periodic boundary 
condition along $\mathbf{a_1}$ is there, so that $k_1$ acts as a 
good quantum number like before. The resulting strip has zigzag left and 
right edges. 
 
Here, a finite strip of stuffed Honeycomb lattice is considered which has 
$N=100$ cells, {\em i.e.}, 300 sites along $\mathbf{a_2}$ direction. 
The $3N\times 3N$ Hamiltonian has been constructed, which is  
a function of $k_1$ by taking partial Fourier transformation. 
Diagonalizing that Hamiltonian, 
the dispersion relation is obtained 
for $t_2=0.8,\,t_3=0.5,\,\epsilon=3.0$, shown in Fig \ref{edge} (a), 
which reveals that the pattern of edge states supports the 
pattern of Chern numbers $(0,-1,1)$ for the relevant topological phase 
of the system. 
Similarly, for other topological phase, $C_n=(0,2,-2)$, a  
honeycomb structure composed of $N= 200$ cells along 
$\mathbf{a_2}$ axis is considered. 
The dispersion relation for $t_2=6.0,\,t_3=2.0,\,\epsilon=6.0$ 
is shown in Fig \ref{edge} (b). 
To maintain a rich clarity in the figure, only two upper bands 
containing the edge states are shown in the second case. 
The edge states are indeed localized in either left (red curves) 
or right (blue curves) edge of the finite lattice, 
as shown in the lower panels of Fig \ref{edge}. 
Chiral nature of these edge states are also confirmed
from the figures, since 
the right-going (left-going) states are always localized in the 
right (left) edge.

As the lowest band always carries 
zero Chern number, no edge states are found to exist in the lower band gap 
for both phases. 
Two pairs of edge states are found in the upper band gap for $C_n=(0,2,-2)$,    
 in contrast to one pair of that 
in the upper band gap for $C_n=(0,-1,1)$ phase. 
Those results are in accordance with the `bulk-boundary 
correspondence' rule which states that:
sum of the Chern number upto the $i$-th band,  
\(\nu_i = \sum_{j\leqslant i}C_j \), is equal to the 
number of pair of edge states in the gap \cite{Mook}. Thus the 
values of the Chern numbers can be recovered from the 
edge state pattern itself.


 \section{Summary and Discussion}
 \label{summary}
A three-band tight-binding model on the stuffed honeycomb lattice has been proposed 
where an additional phase associated with the NNN hopping terms is found to break the 
TRS by yielding a Chern insulating phase  
in the presence of NN hopping.  Although the same phase 
coupled with the NN hopping breaks TRS, it does not give rise to any nontrivial 
topological phase. 
The phase is chosen in such a way that the net flux of gauge field 
per unit cell vanishes. 
This system exhibits two different 
TI phases, Chern insulating and 
 semi-metallic with nonzero Chern numbers in the parameter regimes. 
Also, this one undergoes quantum phase transition between two topological phases 
driven by the hopping parameters. 
Hall conductance exhibits prominent IQHE plateaus. 
The emergence of topologically protected chiral edge states in a strip configuration with open 
boundary condition is also found when $C_n \neq 0$. 

This study reveals the fact that additional triangular sublattice 
leads to the generation of one additional topological phase in the 
resulting stuffed honeycomb lattice. It is established that 
topological phases could be induced within the trivial systems 
by means of introducing phase dependent hopping terms \cite{Haldane}. 
So, more additional topological phases may be obtained 
by choosing those phases and hoppings in different ways. Topological phases 
with higher Chern numbers could be realized within a  
system by means of either exposing the systems to polarized light \cite{Arghya} 
or invoking distant-neighbor 
hopping terms \cite{Piechon,Moumita}. Addition of extra sublattice may 
pave another route for the engineering of new topological phases 
wherever it is applicable. 
Thus, a desired topological phase may be realized in any 
system by combining those procedures along with the 
incorporation of additional sublattice. 
In this context, a tight-binding model is formulated on a body-centered 
stuffed square lattice which is found to harbour new topological 
phase. A brief description
of the system as well as its topological
properties is available in Appendix \ref{Square}.

The picture of stuffed honeycomb lattice is brought into light 
in the context of AFM compound, LiZn$_{2}$Mo$_{3}$O$_8$ \cite{Flint}. 
In this material the molecular cluster Mo$_{3}$O$_8$ forms 
a triangular lattice, however, its spin-liquid property is 
explained in terms of an effective spin-1/2 Heisenberg model 
built on the stuffed honeycomb lattice. 
Likewise, magnetic properties of the partially frustrated triangular 
AFM material, RbFeBr$_3$, was explained before in terms of 
distorted triangular lattice, 
which is essentially the stuffed honeycomb lattice \cite{Adachi}.
However, no material is available at this moment where this electronic 
model could be realized for the verification of its topological 
properties. 

 \section{ACKNOWLEDGMENTS}
AS acknowledges the CSIR fellowship, no. 09/096(0934)  (2018), India.  
AKG acknowledges BRNS-sanctioned 
research project, no. 37(3)/14/16/2015, India. 
\appendix
 \section{The Gell-Mann matrices}
 \label{gm}
 The Gell-Mann matrices are a set of eight 
traceless $3\times3$ linearly independent Hermitian matrices 
spanning the Lie algebra of the SU(3) group.
 They are given below.
 \begin{flalign}
 \lambda_1=\pmat{0&1&0\\1&0&0\\0&0&0}, \lambda_2=\pmat{0&-i&0\\i&0&0\\0&0&0},
 \lambda_3=\pmat{1&0&0\\0&-1&0\\0&0&0}, 
 \nonumber\\
 \lambda_4=\pmat{0&0&1\\0&0&0\\1&0&0}, \lambda_5=\pmat{0&0&-i\\0&0&0\\i&0&0}, \lambda_6=\pmat{0&0&0\\0&0&1\\0&1&0},
 \nonumber\\
 \lambda_7=\pmat{0&0&0\\0&0&-i\\0&i&0},
 \lambda_8=\frac{1}{\sqrt{3}}\pmat{1&0&0\\0&1&0\\0&0&-2}. 
 \end{flalign}
 
 The off-diagonal upper triangular terms of $H(\textbf{k})$ are written as
\begin{equation}
 \begin{aligned}
  h_1-ih_2&= t_1 e^{i \sum_j \mathbf{k}\cdot\mathbf{\delta_j}} \\
  h_4-ih_5&= t_2 e^{i \sum_j \mathbf{k}\cdot\mathbf{\delta_j}} \\
  h_6-ih_7&= t_2 e^{-i \sum_j \mathbf{k}\cdot\mathbf{\delta_j}}
 \end{aligned}
 \end{equation}
where $ \mathbf{k}=(k_x,k_y) $

\section{Stuffed Square Lattice}
\label{Square}
\begin{figure*}
\centering
 \psfrag{Kx}{$k_x$}
 \psfrag{Ky}{$k_y$}
\psfrag{c2}{\text{\scriptsize{\bf \color {white} {$C_n=2$}}}}
\psfrag{c-4}{\text{\scriptsize{\bf \color {white} {$C_n=-4$}}}}
\includegraphics[width=3cm,height=3.2cm,trim={0.0cm 0.0cm 0.0cm 0.0cm}]{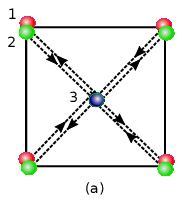}
\includegraphics[width=6cm,height=6cm,trim={1.0cm 0.0cm 2.0cm 0.0cm}]{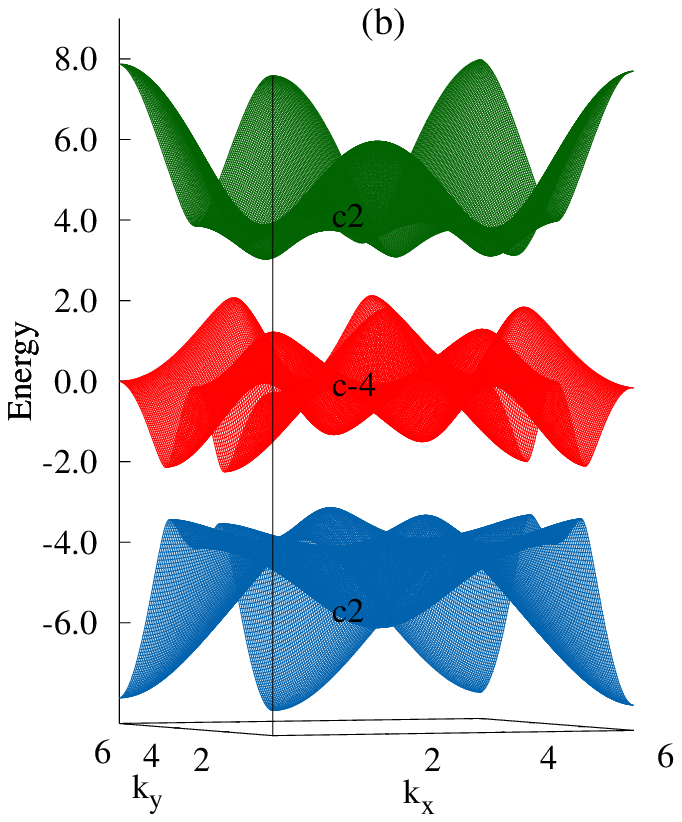}
\includegraphics[width=7cm,height=5cm,trim={0.0cm 0.0cm 0.0cm 0.0cm}]{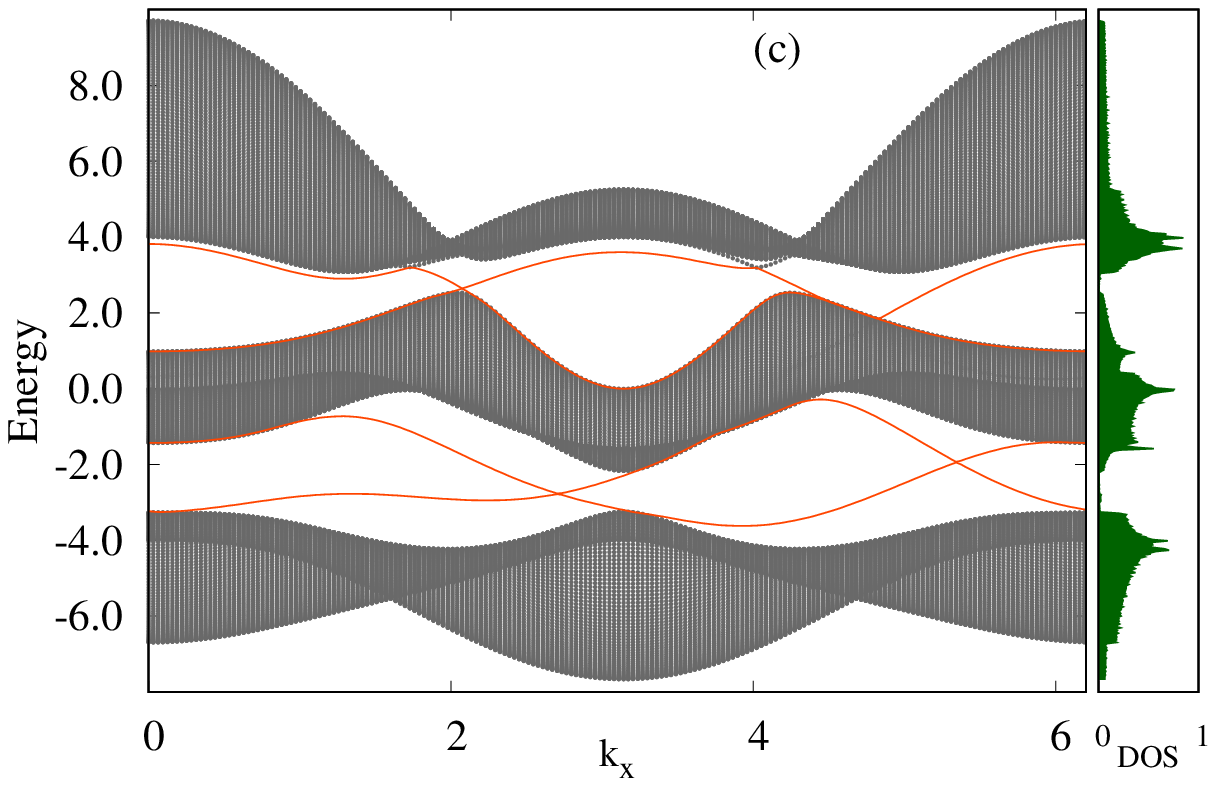}
\caption{ (color online) (a) Schematic diagram of a unit cell of the stuffed square lattice.
  Three sublattices 1,2,3 are shown as red, green and blue spheres, respectively.
   Arrows in the
  NN bond between (1, 2) and 3 show the direction of spin orbit coupling.
  (b) Energy spectrum for $t_1=1.0,t_2=1.2,\lambda_{\rm SOC}=0.8$. Chern numbers of respective
  bands are specified.
  (c) Edge states of the finite system for the same
  parameters and 100 unit cells along $y$-direction. Gapless edge states are
  shown by solid red lines. Density of states is shown in the side panel.
}
 \label{square}
 \end{figure*}

In order to justify our claim that incorporation of additional 
sublattice may induce new topological phase, we consider the 
example of a particular stuffed square lattice. 
In this body-centered square lattice, a two-orbital (denoted by $I=1,2$) 
square sub-lattice incorporates another 
single-orbital (denoted by $I=3$) square sub-lattice 
in its centers which makes the system a three-band
model, as shown in Fig. \ref{square}(a). However, in this case, 
the system is driven towards 
non-trivial topology phase as soon as the SOC 
between (1, 2) and 3 is invoked. 
The Hamiltonian of
this tight binding model on this lattice can be written as
\bea
H=H_{\rm NNN}+H_{\rm NN}+H_{\rm SOC}
\eea
where
\bea
H_{\rm NNN}&=&-\!\!\!\!\!\!\sum_{\langle\langle jk\rangle\rangle,I=1,2}\!\!\!\!\!\left(t_{jk} + (-1)^{I+1}\,t_1\right)a_{jI}^{\dagger}a_{kI},\\
H_{\rm NN}&=&  -\sum_{\langle jk\rangle}t_2\,a_{jI}^{\dagger}a_{kI},\\
H_{\rm SOC}&=& -i\,\lambda_{\rm SOC}\sum_{\langle jk\rangle}\nu_{jk}\,a_{jI}^{\dagger}\sigma_{z}a_{kI}.
\eea
Here, $t_1$ is the NNN hopping parameter between sublattice 1 and 2 while $t_2$ is the NN hopping parameter
between sublattice (1, 2) and 3. $\lambda_{\rm SOC}$ is the strength of SOC between NN pairs. 
The term $\nu_{jk}=+1 \,(-1)$, when the hopping takes place 
along (opposite to) the direction of the arrow as shown in fig \ref{square}(a).
 $a_j$ ($a_j^{\dagger}$) is the annihilation (creation) operator of 
electron at site $j$, such that 
$a_j=(a_{j\uparrow},a_{j\downarrow})^{T}$, where, $a_{j\uparrow}$ ($a_{j\downarrow}$) 
is the annihilation operator of the electron with 
up (down) spin.
$t_{jk}=+t_1(-t_1)$ for the horizontal (vertical) NNN bond.

After applying Fourier transformation, the resulting Hamiltonian, 
$H_{\mathbf{k}}$, comprises two uncoupled diagonal blocks
corresponding to up and down spins, those are related by 
time-reversal symmetry, {\em i. e.},
$H_{\mathbf{k}}^{\uparrow}=H_{-\mathbf{k}}^{\downarrow\ast}$. Now it is sufficient to 
consider either up- or down-spin sector for the 
calculation of energy spectrum and topological properties. It is noted that,
the up- and down-spin sectors break TRS separately. 
So, we can treat Chern number as the topological invariant following the 
previous formulation as described in the main text.
If we denote the Chern number for $n$-th band for up-spin sector as $C_n^{\uparrow}$ and down-spin sector as
$C_n^{\downarrow}$, then $C_n^{\downarrow}=-C_n^{\uparrow}$ 
and spin Chern number for the whole $n$-th band 
can be defined as $C_n^{s}=C_n^{\uparrow}-C_n^{\downarrow}=2C_n^{\uparrow}$. 
In the following, we
restrict to the spin up sector only and omit the spin symbol for convenience. 
So,
\be H\left(\textbf{k}\right)=\sum_k {\psi_k}^{\dagger} h\left(\textbf{k}
\right) \psi_k, \ee
where $\textbf{k} =(k_x,k_y)$ and $\psi_{\textbf{k}} = \left(c_{\textbf{k}1},c_{\textbf{k}2},c_{\textbf{k}3}\right) $
is a three-component spinor. $h(\textbf{k})$ can be expressed as
\be  h\left(\textbf{k}\right) = \sum_i h_i\, \lambda_i, \ee
 with
 \begin{equation}
 \begin{aligned}
 h_1&=2t_1\left(\cos(k_x)-\cos(k_y)\right), \\
 h_2&=0=h_8, \\
 h_3&=2t_1\left(\cos(k_x)+\cos(k_y)\right), \\
 h_4&=4t_2\cos(k_x/2)\cos(k_y/2) = h_6, \\
 h_5&=-4\,\lambda_{\rm SOC}\,\sin(k_x/2)\sin(k_y/2) = -h_7, \\
 \end{aligned}
 \end{equation}

When $\lambda_{\rm SOC}=0.0$, the three bands touch each other and so 
the Chern numbers are ill-defined.
As soon as $\lambda_{\rm SOC}$ becomes non-zero, gaps open up between
 the bands and this time those bands are characterized
 by definite Chern numbers, $C_n=(2,-4,2)$. 
Fig. \ref{square} (b) shows the band structure for
 $t_1=1$, $t_2=1.2$ and $\lambda_{\rm SOC}=0.8$.

For the calculation of edge states, we consider a finite strip of stuffed square lattice
consisting of 100 cells {\em i. e.}, 300 sites along $y$-direction. By diagonilizing the resulting
Hamiltonian as a function of good quantum number $k_x$, the dispersion of edge states is
obtained, which is shown in Fig. \ref{square}(c). 
Two pair of edge states are found in both the band-gaps, 
which is in accordance with the `bulk-boundary correspondence' rule. 
DOS is shown in the side panel of the associated figure. 
Therefore, we put forward another example through which 
we successfully demonstrate the emergence of new topological phase 
with the addition of 
extra sublattice where the tight-bindig model on the parent lattice 
was topologically trivial.


\begin{thebibliography}{99}
\bibitem{TKNN}  Thouless D. J., Kohomoto M., Nightingale P. and den Nijs M.,  
Phys. Rev. Lett. {\bf 49}, 405 (1982).
\bibitem{Kane1}  Kane C. L. and  Mele E. J., 
Phys. Rev. Lett. {\bf 95}, 146802 (2005).
\bibitem{SSH}  Heeger A. J., Kivelson S., Schrieffer J. R. and  Su W. P., 
Rev. Mod. Phys. {\bf 60}, 781 (1988). 
\bibitem{Haldane} Haldane F. D. M.,  
 Phys. Rev. Lett. {\bf 61}, 2015 (1988). 
\bibitem{Hatsugai1}  Hatsugai Y., 
Phys. Rev. Lett. {\bf 71}, 3697 (1993).
\bibitem{Hatsugai2}  Hatsugai Y., Phys. Rev. B {\bf 48}, 11851 (1993). 
\bibitem{Kane2}  Hasan M. Z. and Kane C. L., Rev. Mod. Phys. {\bf 82}, 3045 (2010).
\bibitem{Das} Das T. A, arXiv:1604.07546
\bibitem{Franz} Weeks  C. and Franz M., 
Phys. Rev. B {\bf 82}, 085310 (2010).
\bibitem{Wen} Tang E., Mei J. W. and  Wen X. G., 
Phys. Rev. Lett. {\bf 106}, 236802 (2011). 
\bibitem{Das-sharma1} Sun K., Gu Z., Katsura  H. and Das Sarma S., 
  Phys. Rev. Lett. {\bf 106}, 236803 (2011).
\bibitem{Kargarian} Kargarian M. and Fiete G. A., 
Phys. Rev. B {\bf 82}, 085106 (2010).
\bibitem{Gong2} Liu X.P., Chen W. C., Wang Y. F. and  Gong C. D.,  J. Phys.: 
Condens. Matter {\bf 25}, 305602 (2013). 
\bibitem{Wang}  Wang F. and  Ran Y.,  
Phys. Rev. B {\bf 84}, 241103(R) (2011).
\bibitem{Chen} Chen W. C., Liu R., Wang Y. F. and  Gong C. D.,  
Phys. Rev. B {\bf 86}, 085311 (2012).
\bibitem{Das-sharma2} Yang S., Gu Z., Sun K. and  Das Sarma S., 
 Phys. Rev. B {\bf 86}, 241112(R) (2012).
\bibitem{Wu} Wu C., Phys. Rev. Lett. {\bf 101}, 186807 (2008).
\bibitem{Weiss} Eckardt A., Weiss C. and Holthaus M.,  
 Phys. Rev. Lett. {\bf 95}, 260404 (2005).
\bibitem{Spielman} Jim\'enez-Garc\'ia K. {\em et. al.},
Phys. Rev. Lett. {\bf 108}, 225303 (2012).
\bibitem{Aidelsburger} Aidelsburger M. {\em et. al.}, Nature Phys. {\bf 11}, 162 (2015).
\bibitem{Bergholtz} Trescher M. and Bergholtz E. J., 
Phys. Rev. B {\bf 86}, 241111(R) (2012).
\bibitem{Gong1} Liu R., Chen W.C., Wang Y. F. and Gong C. D.,  
 J. Phys.: Condens. Matter {\bf 24}, 305602 (2012).
\bibitem{Beugeling} Beugeling W., Everts  J. C. and  Morais Smith C.,  
Phys. Rev. B {\bf 86}, 195129 (2012).
\bibitem{Sahoo} Sahoo J., Kochkov D., Clark  B. K. and Flint R.,  
 Phys. Rev. B {\bf 98}, 134419 (2018).
\bibitem{Nagaosa}Nagaosa N. {\em et. al.}, Rev. Mod. Phys. {\bf 82}, 1539 (2010).
\bibitem{Das2} Ray S., Ghatak A. and Das T., Phys. Rev. B {\bf 95}, 165425 (2017).   
\bibitem{Suzuki} Fukui T., Hatsugai  Y. and  Suzuki H.,  
J. Phys. Soc. Jpn. {\bf 74}, 1674 (2005).
\bibitem{Wang1}  Wang Y-F., Hong Y., Gong C-D., and Sheng D. N.,   
Phys. Rev. B {\bf 86}, 201101(R) (2012).
\bibitem{Mook} Mook A., Henk J. and Mertig I., Phys. Rev. B {\bf 90}, 024412 (2014).
\bibitem{Arghya} Sil A. and Ghosh A. K., J. Phys.: Condens. Matter {\bf 31}, 245601 (2019).
\bibitem{Piechon} Sticlet D. and Pi\'echon F., Phys. Rev. B {\bf 87}, 115402 (2013).
\bibitem{Moumita} Deb M. and Ghosh A. K., J. Phys.: Condens. Matter {\bf 31}, 345601 (2019).
\bibitem{Flint} Flint R. and Lee P. A., Phys. Rev. Lett. {\bf 111}, 217201 (2013). 
\bibitem{Adachi} Adachi K. {\em et al.}, 
J. Phys. Soc. Jpn. {\bf 52}, 2202 (1983). 
\end{thebibliography}
\end{document}